\documentclass[draftcls, onecolumn, 11pt]{IEEEtran}
\usepackage{cite}
\usepackage{setspace}
\usepackage[top=1in, bottom=1in, left=1.25in, right=1.25in]{geometry}
\usepackage{color,yfonts}
\usepackage[pdftex]{graphicx}
\graphicspath{{fig/}{jpeg/}}
\usepackage[cmex10]{amsmath}
\usepackage{amssymb}
\usepackage{epstopdf,epsfig}
\usepackage{multirow}
\usepackage{subfigure} 
\usepackage{algorithm}
\usepackage{tikz}
\usetikzlibrary{shapes,arrows,matrix}
\usetikzlibrary{decorations.pathmorphing} 
\usetikzlibrary{fit}                    
\usetikzlibrary{backgrounds}    
\usepackage{verbatim}
\usepackage{algpseudocode}
\usepackage{amsmath,amssymb,lipsum}
\usepackage{hyperref}
\usepackage{comment}
\usepackage{graphicx}
\usepackage{mathtools,enumerate}
\usepackage{multicol,lipsum}

\usepackage{stix}

\input{mysymbol.sty}
\renewcommand{\theenumi}{\roman{enumi}}%

\newcommand{\ayche}{\mathscr{h}}

\newcommand{\overbar}[2][3]{{}\mkern#1mu\overline{\mkern-#1mu#2}}
\newcommand{\INDSTATE}[1][1]{\State\hspace{3mm}}
\newcommand{\INDSTATED}[1][1]{\State\hspace{6mm}}

\usepackage{theorem}

\theoremstyle{definition}

\def \x {{\mathbf{x}}}

 \providecommand{\norm}[1]{\left\|#1\right\|}

  \tikzstyle{agent}=[circle,
  thick,
  minimum size=.5cm,
  draw=blue!80,
  fill=blue!20]

  \tikzstyle{neighbor}=[circle,
  thick,
  minimum size=.5cm,
  draw=cyan!85!black,
  fill=cyan!40,
  decorate,
  decoration={random steps,
      segment length=2pt,
      amplitude=2pt}]
  
  \tikzstyle{local_nat}=[rectangle,
  thick,
  minimum size=.5cm,
  draw=red!80,
  fill=red!20]
  
  \tikzstyle{glob_nat}=[rectangle,
  thick,
  minimum size=.5cm,
  draw=red!100,
  fill=red!40]      
  
  \tikzstyle{background}=[rectangle,
  fill=gray!10,
  inner sep=0.2cm,
  rounded corners=5mm]
  
  \tikzstyle{background2}=[rectangle,
  fill=green!20,
  inner sep=0.2cm,
  rounded corners=5mm]
{\tiny }
    \title{\vspace{-0cm}{Optimally Compressed \\ Nonparametric Online Learning}}

\author{Alec~Koppel,~\IEEEmembership{Member,~IEEE,}
	Amrit~Singh~Bedi,~\IEEEmembership{Student Member,~IEEE,}
	Ketan~Rajawat,~\IEEEmembership{Member,~IEEE} and Brian~M.~Sadler,~\IEEEmembership{Fellow,~IEEE} \thanks{ A. Koppel is a  Research Scientist with the U.S. Army Research Laboratory in the Computational and Information Sciences Directorate (CISD) since Sep. 2017. His research focuses on optimization and learning for autonomy.

A. S. Bedi is a Postdoctoral Associate with the U.S. Army Research Laboratory in CISD since Jan. 2019.  His research interests include stochastic and time-varying optimization for networked systems as well as theoretical machine learning.

K. Rajawat is an Assistant Professor with the Department of Electrical Engineering, Indian Institute
of Technology Kanpur, Kanpur, India.  He is currently an Associate Editor for the IEEE Communication Letters. His
current research focuses on distributed algorithms, online learning, and networked control systems.
 
B. M. Sadler is the U.S. Army Senior Scientist for Intelligent Systems, and a Fellow with the Army Research Laboratory, Adelphi, MD, USA. He was an IEEE Signal Processing Society Distinguished Lecturer for 2017-2018, an Associate Editor for the IEEE Transactions on Signal Processing, IEEE Signal Processing Letters, and EURASIP Signal Processing. His research interests include information science and networked autonomous intelligent systems. }}
\begin{document}
\maketitle
\vspace{-10mm}
\begin{abstract}%
%
 Batch training of machine learning models based on neural networks is now well established, whereas to date streaming methods are largely based on linear models. To go beyond linear in the online setting, nonparametric methods are of interest due to their universality and ability to stably incorporate new information via convexity or Bayes' Rule. Unfortunately, when used online, nonparametric methods suffer a ``curse of dimensionality" which precludes their use: their complexity scales at least with the time index.  We survey online compression tools which bring their memory under control and attain approximate convergence. The asymptotic bias depends on a compression parameter that trades off memory and accuracy. Further, the applications to robotics, communications, economics, and power are discussed, as well as extensions to multi-agent systems.
%
%
%
\end{abstract}

\vspace{-2mm}
\section{Introduction}\label{sec:intro}
\subsection{Motivation}
%
Modern machine learning is driven in large part by training nonlinear statistical models on huge data sets held in cloud storage via stochastic optimization methods.
 This recipe has yielded advances across fields such as vision, acoustics, speech, and countless others. Of course, universal approximation theorems \cite{tikhomirov1991representation}, the underpinning for the predictive capability of methods such as deep neural networks and nonparametric methods, have been known for some time. 
 %
%
However, when accumulating enough data a priori is difficult or if the observations arise from a non-stationary process, as in robotics \cite{thrun2005probabilistic}, energy \cite{atzeni2013demand}, and communications \cite{ribeiro2010ergodic}, \emph{online training} is of interest. Specifically, for problems with dynamics, one would like the predictive power of universality while stably and quickly adapting to new data.

%
Bayesian and nonparametric methods \cite{ghosal2017fundamentals} meet these specifications in the sense that they possess the universal approximation capability and may stably adapt to new data.
In particular, tools under this umbrella such as kernel regression \cite{hofmann2008kernel}, Gaussian Processes, and particle filters/Monte Carlo methods \cite{djuric2003particle} can stably (in a Lyapunov sense) incorporate new information as it arrives via functional variants of convex optimization or probabilistic updates via Bayes' Rule. These facts motivate applying Bayesian and nonparametric methods to streaming problems. 

Alas, Bayesian and nonparametric methods are limited by the curse of dimensionality. Typically, these tools require storing a probability density estimate that retains all past training examples together with a vector of weights that denotes the importance of each sample. Moreover, as time becomes large, their complexity approaches infinity. This bottleneck has led to a variety of approximation schemes, both offline and online, to ensure memory is under control. Typically one fixes some memory budget, and projects all additional training examples/particles onto the likelihood of the density estimate spanned by current points \cite{wang2012breaking}. Alternatives include memory-efficient approximations of the associated functional \cite{williams2001using} or probabilistic model \cite{rahimi2009weighted}.

\subsection{Significance}
Until recently, a significant gap in the literature of memory reduction design for Bayesian and nonparametric methods existed: how to provide a tunable tradeoff between \emph{statistical consistency}\footnote{Consistency is the statistical version of optimality, and means that an empirical estimator converges to its population analogue, i.e., is asymptotically unbiased.} and memory? Recently, the perception that nonparametric methods do not scale to streaming settings has been upended \cite{koppel2019parsimonious,bedi2019nonparametric}. The key insight of these methods is that one may fix an error-neighborhood around the current density estimate, and project it onto a nearby subspace with lower memory, thus sparsifying the sample path \emph{online} and allowing the  complexity of the density representation to be flexible and problem-dependent \cite{elvira2017adapting}.  That is, the radius of the error neighborhood, which we henceforth call the \textit{compression budget}, determines the asymptotic radius of convergence, i.e., bias, as well as the complexity of the resulting density function.  
 
{The performance guarantees of this approach are reminiscent of the rate distortion theoretic results in the context of lossy data compression.
Recall that the rate distortion theory characterizes the number of bits
required to approximately represent a signal while allowing for a
pre-specified distortion. In a similar vein, the compression budget
$\epsilon$ determines the number of data points required to attain a
nonparametric statistical model that is $\epsilon$ -statistically
consistent. In other words, the model order plays the role of the number
of bits that quantify the amount of information contained in the
nonparametric model, while the compression budget is analogous to the
maximum allowable distortion. The analogy continues to hold across the
family of nonparametric models such as kernel density estimates \cite{ghosal2017fundamentals}, supervised learning \cite{koppel2019parsimonious}, Gaussian processes \cite{koppelconsistent}. In this review, we focus on supervised learning.}

  \begin{figure} 
  \centering
  \includegraphics[scale=0.4]{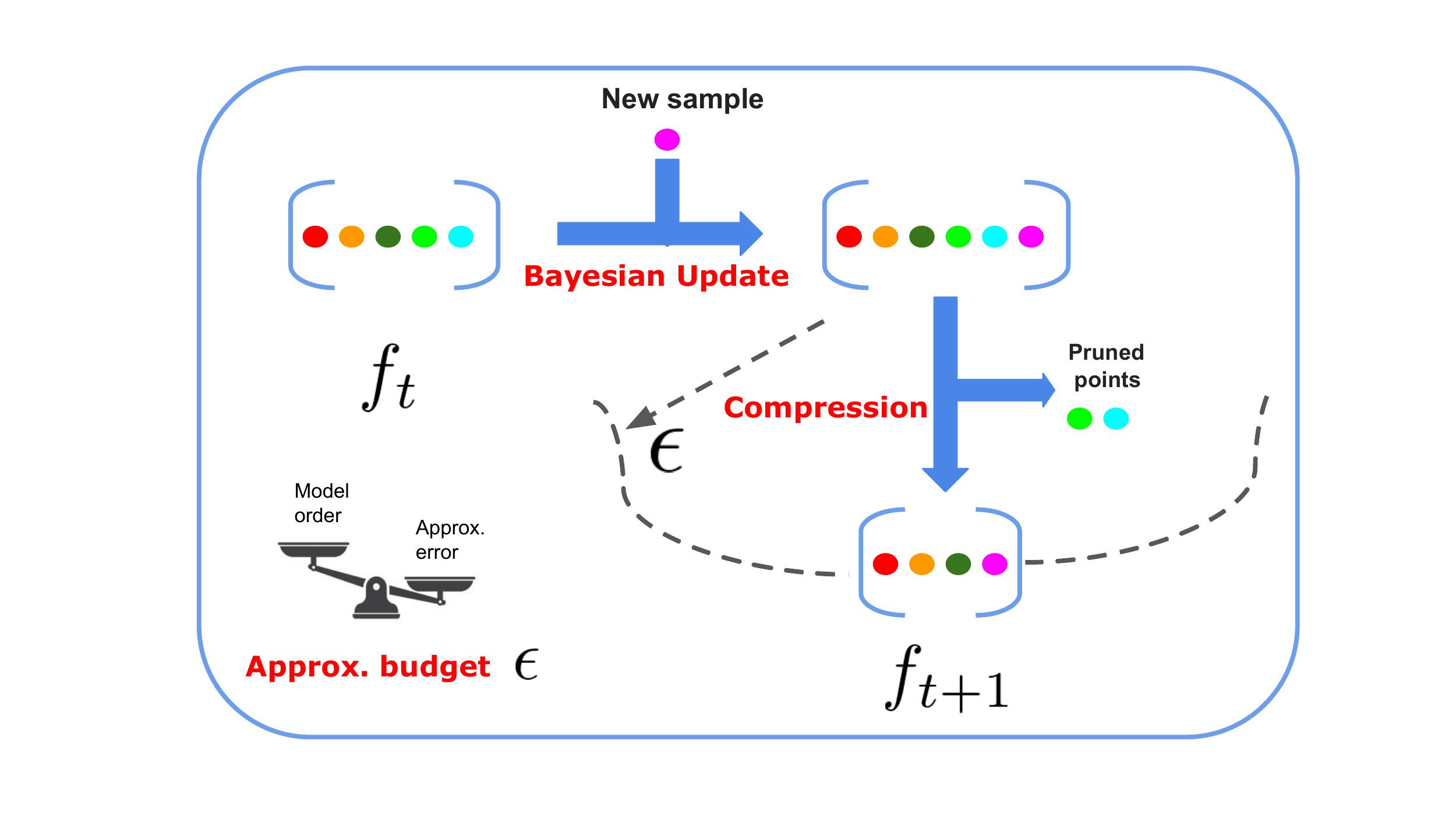} 
  \vspace{-8mm}\caption{Generalized Projection scheme for Bayesian and nonparametric methods in the online setting. A new sample arrives and is incorporated as either functional stochastic gradient method, maximum a posteriori estimation, or Monte Carlo particle generation. Rather than allow the complexity to grow ad infinitum, the current statistical model is projected a lower dimensional subspace via greedy compression which is at most $\epsilon$-far away according to some metric. The compression parameter $\epsilon$ trades off statistical consistency and memory requirements.}\vspace{-6mm}
  \end{figure}
\subsection{Impact}
Proposed approaches allow stable and memory-efficient online training of universal statistical models, a foundation upon which many theoretical and practical advances may be built. 
%
For instance, in autonomous control of robotic platforms, it is well-known that optimal control based on physical models \cite{garcia1989model} is limited to the domain for which the model remains valid. When used outside this domain, a gap emerges between an a priori hypothesized physical model and the ground truth. Alternatively, such problems can be cast within the rubric of sequential learning where one seeks to regress on the model mismatch. Initial efforts toward this end have appeared recently \cite{koller2018learning}, but enhancements are possible, for instance, by increasing the descriptive capacity of the class of learned models and the adaptivity of autonomous behaviors to changing environments.

 A similar problem arises in communications systems, where nonlinear adaptive filters can be used to mitigate signal saturation in amplifiers and detectors. The current techniques are limited to linearity such as for equalization in receivers, noise cancellation, beamforming in radar, and sonar systems.
 This limitation is because current communications channel state estimators \cite{ribeiro2010ergodic}, modulators, and localizers need to adapt with low latency, and typically only linear methods meet the requirement \cite{kaleh1994joint}. However, nonparametric and Bayesian methods would enable identification and tracking of modulation type from symbol estimates that exhibit more complex interaction effects, channel state estimation in the presence of nonlinearities arising from the environment, and localization problems where the entire distribution over possible source locations is of interest, due to, e.g., desire for confidence guarantees.
  
%
 
On the theory side, we note that efforts to develop optimization tools for multi-agent online learning have been mostly restricted to cases where agents learn linear statistical models \cite{nedic2009distributed}, which exclude the state of the art machine learning methods. However, since kernel learning may be formulated as a stochastic convex problem over a function space, standard strategies, i.e., distributed gradient \cite{koppel2018decentralized} and primal-dual methods \cite{pradhan2018exact} may be derived. These schemes allow individual agents to learn in a decentralized online manner a memory-efficient nonlinear interpolator as good as a centralized clairvoyant agent with all information in advance. Many enhancements are possible via recent advances in decentralized optimization \cite{shi2015extra}, such as those which relax conditions on the network \cite{nedic2015distributed} and amount of communications \cite{wan2009event}. 

Apart from multi-agent systems, the general approach of sparsifying as much as possible while ensuring a descent property holds, may be applied to other nonparametric statistical tools such as Gaussian processes and Monte Carlo Methods \cite{koppelconsistent,elvira2017adapting} by varying the ambient space and compression metric. Due to the need for brevity, we defer such a discussion to future work.  

The rest of the paper is organized as follows. In Section \ref{sec:kernels}, we formulate the problem of supervised online training with kernels, and define memory-affordable kernel regression algorithms for solving them in \ref{sec:polk} and extend this framework to online risk-aware learning (Section \ref{sec:colk}). In Section \ref{sec:decentralized}, we spotlight online nonparametric learning methodologies in decentralized settings. In Section \ref{sec:conclusion}, we conclude with a discussion of implications and open problems.


\section{Online Supervised Learning with Kernels}\label{sec:kernels}
We begin by formalizing the problem of supervised learning as expected risk minimization  (ERM), which is the foundation of filtering, prediction, and classification. Then, we detail how the problem specializes when the estimator admits a kernel parameterization.
In ERM,  we seek a function that minimizes a loss quantifying the merit of a statistical model averaged over data. Each point in the data set {$\{(\bbx_n, y_n)\}_{n=1,\dots}$} constitutes  an input-output pair $(\bbx_n, y_n)$ which is an i.i.d. realization from the stationary distribution of random pair $(\bbx, y) \in \ccalX \times \ccalY$ with $\ccalX\subset \reals^p$. We consider the problems where samples arrive sequentially, which is applicable to signal processing, communication,  visual perception, and many other applications.

Hereafter, we quantify the quality of an estimator function $f\in\mathcal{H}$ by a convex loss function $\ell:\ccalH \times \ccalX \times \ccalY \rightarrow \reals$, where $\ccalH$ is a hypothesized function class. In particular, if we evaluate $f$ at feature vector $\bbx$, its merit is encapsulated by $\ell(f(\bbx), y)$. Then, we would like to select $f$ to have optimal model fitness on average over data, i.e., to minimize the statistical loss $L(f) : = \mbE_{\bbx, y}{[ \ell(f(\bbx), y)]}$.
 We focus on minimizing the regularized loss $R(f) : = \argmin_{f \in \ccalH} L(f) +(\lambda/2)\|f \|^2_{\ccalH}$ as\vspace{-1mm}
\begin{align}\label{eq:kernel_stoch_opt}
f^\star=\argmin_{f \in \ccalH} R(f) :& = \argmin_{f \in \ccalH}\mbE_{\bbx, y}\Big[ \ell(f\big(\bbx), y\big)\Big] +\frac{\lambda}{2}\|f \|^2_{\ccalH} 
\end{align}
The preceding expression specializes to linear regression when $f(\bbx) = \bbw^T\bbx$ for some $\bbw\in\reals^p$, but rather here we focus on the case that function class $\mathcal{H}$ is a reproducing kernel Hilbert space (RKHS) -- for background, see \cite{hofmann2008kernel}.
An RKHS consists of functions $f:\mathcal{X}\rightarrow\mathcal{Y}$ that have a basis expansion in terms of elements of $\mathcal{H}$ through a kernel $\kappa: \ccalX \times \ccalX \rightarrow \reals$ defined as: \vspace{-1mm}
\begin{align} \label{eq:rkhs_properties}
(i) \  \langle f , \kappa(\bbx, \cdot)) \rangle _{\ccalH} = f(\bbx) \quad \text{for all } \bbx \in \ccalX \; ,
\qquad (ii) \ \ccalH = \overbar{\text{span}\{ \kappa(\bbx , \cdot) \}} \quad\text{for all } \bbx \in \ccalX\; .
\end{align}
where $\langle \cdot , \cdot \rangle_{\ccalH}$ is the Hilbert inner product for $\ccalH$, and the overbar denotes set closure in (ii). The kernel function $\kappa(\cdot,\cdot)$ is henceforth assumed  positive semidefinite: $\kappa(\bbx,\bby)\geq 0$ for all $(\bbx,\bby)\in\mathcal{X}$. Example kernels include the polynomial $\kappa(\bbx,\bby) = \left(\bbx^T\bby+b\right)^c $ and Gaussian $\kappa(\bbx,\bby) = \exp\left\{ -\frac{\lVert \bbx - \bby \rVert_2^2}{2c^2} \right\}$, where $\bbx, \bby \in \ccalX$.

The term \emph{reproducing} comes from replacing $f$ by $\kappa(\bby,\cdot)$ in  \eqref{eq:rkhs_properties}(i) which yields $ \langle \kappa(\bby, \cdot) , \kappa(\bbx, \cdot) \rangle_{\ccalH} = \kappa(\bbx, \bby)$. We note that this reproducing property permits writing the inner product of the feature maps of two distinct vectors $\bbx$ and $\bby$ as the kernel evaluations $\langle \phi(\bbx), \phi(\bby) \rangle_{\ccalH} =\kappa(\bbx, \bby)$. Here, $\phi(\cdot)\in\mathcal{H}$ denotes the feature map of vector $\x\in\mathcal{X}$. The preceding expression, the \emph{kernel trick}, allows us to define arbitrary nonlinear relationships between data without ever computing $\phi$ \cite{hofmann2008kernel}.
Moreover, \eqref{eq:rkhs_properties} (ii) states that any function $f\in\mathcal{H}$ is a linear combination of kernel evaluations at the vectors $\x\in\mathcal{X}$, which for empirical versions of \eqref{eq:kernel_stoch_opt}, implies that the Representer Theorem holds \cite{hofmann2008kernel}\vspace{-1mm}
%
%
\begin{equation}\label{eq:kernel_expansion}
f(\bbx) = \sum_{n=1}^N w_n \kappa(\bbx_n, \bbx)\; .
\end{equation}
Here $\bbw = [w_1, \cdots, w_N]^T \in \reals^N$  a collection of weights and $N$ is the sample size, assuming for the moment that $N<\infty$. We note that the number of terms in the sum \eqref{eq:kernel_expansion}, hereafter referred to as the \emph{model order}, coincides with the sample size, and hence grows unbounded as empirical approximations of \eqref{eq:kernel_stoch_opt} approach their population counterpart. {Eqn. \eqref{eq:kernel_expansion} makes computing the RKHS inner product easy: for $f=\sum_m w_m \kappa(\bbx_m,\cdot)$ and $g=\sum_n w_n \kappa(\bbx_n,\cdot)$, $\langle f, g \rangle = \sum_{m,n} w_m w_n \kappa(\bbx_m, \bbx_n )$. The induced norm is simply $\langle f, f \rangle^{1/2} = \|f\|_{\ccalH}$ }.
%
%
%
%
%
This complexity bottleneck is a manifestation of the \textit{curse of dimensionality} in nonparametric learning. Decades of research has attempted to overcome this issue through the design of memory-reduction techniques, most not guaranteed to yield the minimizers of \eqref{eq:kernel_stoch_opt}. In the following subsection, we detail a stochastic approximation algorithm which explicitly trades off model fitness with memory requirements, 
and provide applications.

\subsection{Trading Off Consistency and Complexity}\label{sec:polk}
In this section, we derive an online algorithm to solve the problem in \eqref{eq:kernel_stoch_opt} through a functional variant of stochastic gradient descent (FSGD) \cite{Kivinen2004}. Then, we detail how memory reduction may be attained through  subspace projections \cite{koppel2019parsimonious}. We culminate by illuminating tradeoffs between memory and consistency both in theory and practice.

We begin by noting that FSGD applied to the statistical loss \eqref{eq:kernel_stoch_opt} when the feasible set is an RKHS \eqref{eq:rkhs_properties} takes the form \cite{Kivinen2004}\vspace{-1mm}
%
%
\begin{align}\label{eq:sgd_hilbert}
f_{t+1}(\cdot) =(1-\eta \lambda ) f_{t} - \eta \nabla_f \ell (f_{t}(\bbx_t),y_t)=(1-\eta \lambda ) f_{t} - \eta \ell'(f_t(\bbx_t),y_t) \kappa(\bbx_t,\cdot) \; ,
\end{align}
where $\eta>0$ is the constant step size. The equality uses the chain rule, the reproducing property of the kernel \eqref{eq:rkhs_properties}(i), and the definition $\frac{\partial \ell (f(\bbx),y)}{\partial f(\bbx) } := \ell ' f(\bbx), y)$ as in \cite{Kivinen2004}. With initialization  $f_0=0$, Representer Theorem \eqref{eq:kernel_expansion} allows us to rewrite \eqref{eq:sgd_hilbert} with dictionary and weight updates as\vspace{-1mm}
\begin{align}\label{eq:param_update} 
\bbX_{t+1} = [\bbX_t, \;\; \bbx_t],\;\;\;\; \bbw_{t+1} = [ (1 - \eta \lambda) \bbw_t , \;\; -\eta\ell'(f_t(\bbx_t),y_t)]  \; .
\end{align}
We define the matrix of training examples $\bbX_t = [\bbx_1,\ \ldots\ ,\bbx_{t-1}]$ as the kernel dictionary, the kernel matrix as $\bbK_{\bbX_t,\bbX_t}\in \reals^{(t-1)\times (t-1)}$, whose entries are kernel evaluations $[\bbK_{\bbX_t, \bbX_t}]_{m,n} =\kappa(\bbx_m, \bbx_n)$, and the empirical kernel map $\boldsymbol{\kappa}_{\bbX_t}(\cdot) = [\kappa(\bbx_1,\cdot) \ldots \kappa(\bbx_{t-1},\cdot)]^T$ as the vector of kernel evaluations.

 %
 \begin{algorithm}[t]
 \caption{Parsimonious Online Learning with Kernels (POLK)} \label{alg:soldd}
 \begin{algorithmic}
 \Require $\{\bbx_t,y_t,\eta_t,\epsilon_t \}_{t=0,1,2,...}$
 \State \textbf{initialize} ${f}_0(\cdot) = 0, \bbD_0 = [], \bbw_0 = []$, i.e. initial dictionary, coefficient vectors are empty
 \For{$t=0,1,2,\ldots$}
 	\State Obtain independent training pair realization $(\bbx_t, y_t)$
 	\State Compute unconstrained functional stochastic gradient step
 	$$\tilde{f}_{t+1}(\cdot) = (1-\eta_t\lambda){f}_t(\cdot) - \eta_t\ell'({f}_t(\bbx_t),y_t)\kappa(\bbx_t,\cdot)$$
 	
 	\State Revise dictionary $\tbD_{t+1} = [\bbD_t,\;\;\bbx_t]$ and weights $\tbw_{t+1} \leftarrow [(1-\eta_t\lambda)\bbw_t,\;\; -\eta_t\ell'({f}_t(\bbx_t),y_t)]$
 	\State Compute sparse function approximation via KOMP
 	$$({f}_{t+1}(\cdot),\bbD_{t+1},\bbw_{t+1}) = \textbf{KOMP}(\tilde{f}_{t+1}(\cdot),\tbD_{t+1},\tbw_{t+1},\epsilon_t)$$
 \EndFor
 \end{algorithmic}
 \end{algorithm}
 
At each time $t$, the new sample $\x_t$ enters into current dictionary $\bbX_t$ to obtain $\bbX_{t+1}$, and hence the \emph{model order} $M_{t+1}=t$, i.e., the number of points in dictionary $\bbX_{t+1}$, tends to $\infty$ as $t\rightarrow\infty$ when data is streaming. Existing strategies for online memory-reduction include dropping past points when weights fall below a threshold \cite{Kivinen2004}, projecting functions onto fixed-size subspaces via spectral criteria \cite{1315946} or the Hilbert norm \cite{williams2001using}, probabilistic kernel approximations \cite{rahimi2009weighted}, and others. A commonality to these methods is a capitulation on convergence in pursuit of memory reduction. In contrast, one may balance these criteria by projecting the FSGD sequence onto subspaces adaptively constructed from past data $\{\bbx_u\}_{u\leq t}$.

{ \noindent \bf Model Order Control via Subspace Projections}
 To control the complexity growth, we propose approximating the function sequence $\{f_t\}$ [cf. \eqref{eq:sgd_hilbert}] by projecting them onto subspaces $\ccalH_{\bbD_t} \subseteq \ccalH$ of dimension $M_t$ spanned by the elements in the dictionary $\bbD_t = [\bbd_1,\ \ldots,\ \bbd_{M_t}] \in \reals^{p \times M_t}$, i.e., $\ccalH_{\bbD_t} = \{f\ :\ f(\cdot) = \sum_{n=1}^{M_t} w_n\kappa(\bbd_n,\cdot) = \bbw^T\boldsymbol{\kappa}_{\bbD_t}(\cdot) \}=\text{span}\{\kappa(\bbd_n, \cdot) \}_{n=1}^{M_t}$, where we denote $[\boldsymbol{\kappa}_{\bbD_t}(\cdot)=\kappa(\bbd_1,\cdot) \ldots \kappa(\bbd_{M_t},\cdot)]$, and $\bbK_{\bbD_t,\bbD_t}$ represents the kernel matrix obtained for dictionary $\bbD_t$. To ensure model parsimony, we enforce the number of elements in $\bbD$ to satisfy $M_t \ll t$.

 To introduce the projection, note that \eqref{eq:sgd_hilbert} is such that $\bbD=\bbX_{t+1}$. Instead, we use dictionary, $\bbD=\bbD_{t+1}$ whose columns are chosen from $\{\bbx_u\}_{u\leq t}$.
 To be specific, we augment \eqref{eq:sgd_hilbert} by projection:
 \begin{align}\label{eq:projection_hat}
\!\! \!{f}_{t+1}(\cdot)\!=\! \argmin_{f \in \ccalH_{\bbD_{t+1}}} \! \Big\lVert  f \!- \!\!
 \Big(\!(1\!-\!\eta_t \lambda) f_t 
 \!-\! \eta_t \nabla_f\ell(f_{t}(\bbx_t),y_t)\! \Big)\!\Big\rVert_{\ccalH}^2
\!\! \!:=\ccalP_{\ccalH_{\bbD_{t+1}}} \!\Big[ \!
 (1\!-\!\eta_t \lambda) f_t 
 \!-\! \eta_t \nabla_f\ell(f_{t}(\bbx_t),y_t) \Big] ,
 \end{align}
 where we denote the subspace $\ccalH_{\bbD_{t+1}}=\text{span}\{ \kappa(\bbd_n, \cdot) \}_{n=1}^{M_{t+1}}$ associated with dictionary $\bbD_{t+1}$, and the right-hand side of \eqref{eq:projection_hat} to define the projection operator. \emph{Parsimonious Online Learning with Kernels} (POLK) (Algorithm \ref{alg:soldd}) projects FSGD onto subspaces $\ccalH_{\bbD_{t+1}}$ stated in \eqref{eq:projection_hat}. Initially the function is $f_0=0$. Then, at each step, given sample $(\bbx_t, y_t)$ and step-size $\eta_t$, we take an \emph{unconstrained} FSGD iterate $\tilde{f}_{t+1}(\cdot) = (1-\eta_t\lambda){f}_t - \eta_t\ell'({f}_t(\bbx_t),y_t)\kappa(\bbx_t,\cdot)$ which admits the parametric representation $\tbD_{t+1}$ and $\tbw_{t+1}$. These parameters are then fed into KOMP (to be defined shortly) with approximation budget $\eps_t$, such that $(f_{t+1}, \bbD_{t+1}, \bbw_{t+1})= \text{KOMP}(\tilde{f}_{t+1},\tilde{\bbD}_{t+1}, \tilde{\bbw}_{t+1},\eps_t)$.

\begin{table*}[]\centering\hfill
\renewcommand{\arraystretch}{1.7}
\caption{Summary of convergence results for diminishing and constant step-size selections.}\vspace{-1mm}
\label{tab2}
\begin{tabular}{llll}
\cline{1-4}  
$\qquad$ & $\qquad$ Diminishing & $\ \quad$ Constant &   $ \ $
\\
\cline{1-4} 
Step-size/Learning rate      & $ \ \qquad\eta_t = \ccalO(1/t)$       &   $\qquad \eta_t = \eta>0$   \\  
Sparse Approximation Budget       &$\ \qquad\eps_t=\eta_t^2$       &  $\qquad\eps=\ccalO(\eta^{3/2})$     \\  
Regularization Condition       & $\ \qquad\eta_t < 1/\lambda$       &  $\qquad\eta < 1/\lambda$     \\  
Convergence Result                & $\ \qquad f_t \rightarrow f^*$ a.s.             & $\qquad\liminf_t \| f_t - f^* \| = \ccalO(\sqrt{\eta})$  a.s. \\ 
Model Order Guarantee               & $\qquad$ None            &  $\qquad$Finite \ \blue{ $\ccalO\left(\big(\frac{\eta}{\eps}\big)^p\right)$ -- see \cite{bedi2019nonstationary}[Lemma 1]}\\ 
\cline{1-4}
\end{tabular}\vspace{-4mm}
\end{table*}

 {\bf \noindent Parameterizing the Projection} The projection may be computed in terms of data and weights. To select dictionary $\bbD_{t+1}$ for each $t$, we use greedy compression, specifically, a destructive variant of \emph{kernel orthogonal matching pursuit} (KOMP) \cite{Vincent2002} with budget $\epsilon$. 
The input function to KOMP is $\tilde{f}$ with model order $\tilde{M}$ parameterized by its kernel dictionary $\tbD\in\reals^{p\times\tilde{M}}$ and coefficient vector $\tbw\in\reals^{\tilde{M}}$. The algorithm outputs $f\in \ccalH$ with a lower model order. We use $\tilde{f}_{t+1}$ to denote the function updated by an un-projected FSGD step, whose coefficients and dictionary are denoted as $\tbw_{t+1}$ and $\tbD_{t+1}=[\bbD_t; \bbx_t]$. 
At each stage, the dictionary element is removed which contributes the least to the Hilbert-norm error $\min_{f\in\ccalH_{\bbD\setminus \{j\}}}\|\tilde{f} - f \|_{\ccalH}$ of the original function $\tilde{f}$, when dictionary $\bbD$ is used. Doing so yields an approximation of $\tilde{f}$ inside an $\epsilon$-neighborhood $\|f - \tilde{f} \|_{\ccalH} \leq \eps_t $. Then, the ``energy" of removed points is re-weighted onto those that remain 
$\bbw_{t+1}=  \bbK_{\bbD_{t+1} \bbD_{t+1}}^{-1} \bbK_{\bbD_{t+1} \tbD_{t+1}} \tbw_{t+1}$.

{While exact characterization of the relationship between $M_t$, the model complexity at time $t$, and the compression budget $\epsilon_t$ is unknown, they have been characterized in an asymptotic worst-case sense in \cite{bedi2019nonstationary}[Lemma 1] an extended version of \cite{1315946}. In particular, ignoring problem constants such as the Lipschitz continuity parameter, it depends on the ratio of the step-size to the compression budget, all raised to the power $p$, where $p$ is the feature-space dimension. This exponential dependence is a typical challenge of nonparametric statistics.}

 {Algorithm \ref{alg:soldd} converges both under diminishing and constant step-size regimes \cite{koppel2019parsimonious} when the learning rate $\eta_t$ satisfies $\eta_t < 1/\lambda$ and is attenuating such that $\sum_t \eta_t = \infty$ and $\sum_t \eta_t^2<\infty$ with approximation budget $\eps_t$ satisfying $\eps_t=\eta_t^2$. Here $\lambda>0$ is the regularization parameter.}
  {Practically speaking, this means that exact convergence requires infinite memory, as $\epsilon_t \rightarrow 0$ means less compression occurs over time and the memory grows.} More interestingly, however, when a constant algorithm step-size $\eta_t = \eta$ is chosen to satisfy $\eta<1/\lambda$, then under constant budget $\eps_t = \eps$ which satisfies $\eps = \ccalO(\eta^{3/2})$, the function sequence which converges to near the optimal $f^*$ [cf. \eqref{eq:kernel_stoch_opt}] and has finite complexity. This tradeoff is summarized in Table \ref{tab2}.

{{\bf Complexity} We note that alternatives to the KOMP-based projection in Algorithm \ref{alg:soldd} are possible based on memory-windowing \cite{Kivinen2004}, nearest neighbor based merging \cite{wang2010online}, projections onto subspaces of fixed sizes \cite{wang2012breaking}, and random feature expansions \cite{rahimi2008random}, among others. These alternatives are lower-complexity, but this lower complexity comes at the cost of directional bias in the update that may only be controlled in expectation. \footnote{{To be specific, memory-reduction of stochastic gradient updates causes bias according to some statistic i.e., the RKHS norm, KL divergence, or spectrum of the kernel matrix. In order to nearly preserve the convergence of an optimization algorithm to which its applied, the memory-reduction method must nearly un-alter its descent properties. POLK and comparable algorithms do so by ensuring the sparsified stochastic gradient is $\epsilon$-close to its dense variant.}}
 See \cite{koppel2019parsimonious}[Remark 1 and Sec. 4.3], where computational effort is spotlighted. The main point is that per-step complexity of computing projections may be made quadratic in the model order multiplied by the number of points that need to be removed, through careful reuse of matrix inverses.}

\begin{figure}[t]
	\centering
	\subfigure[Error rate]{\includegraphics[width=0.5\linewidth,height=5cm]{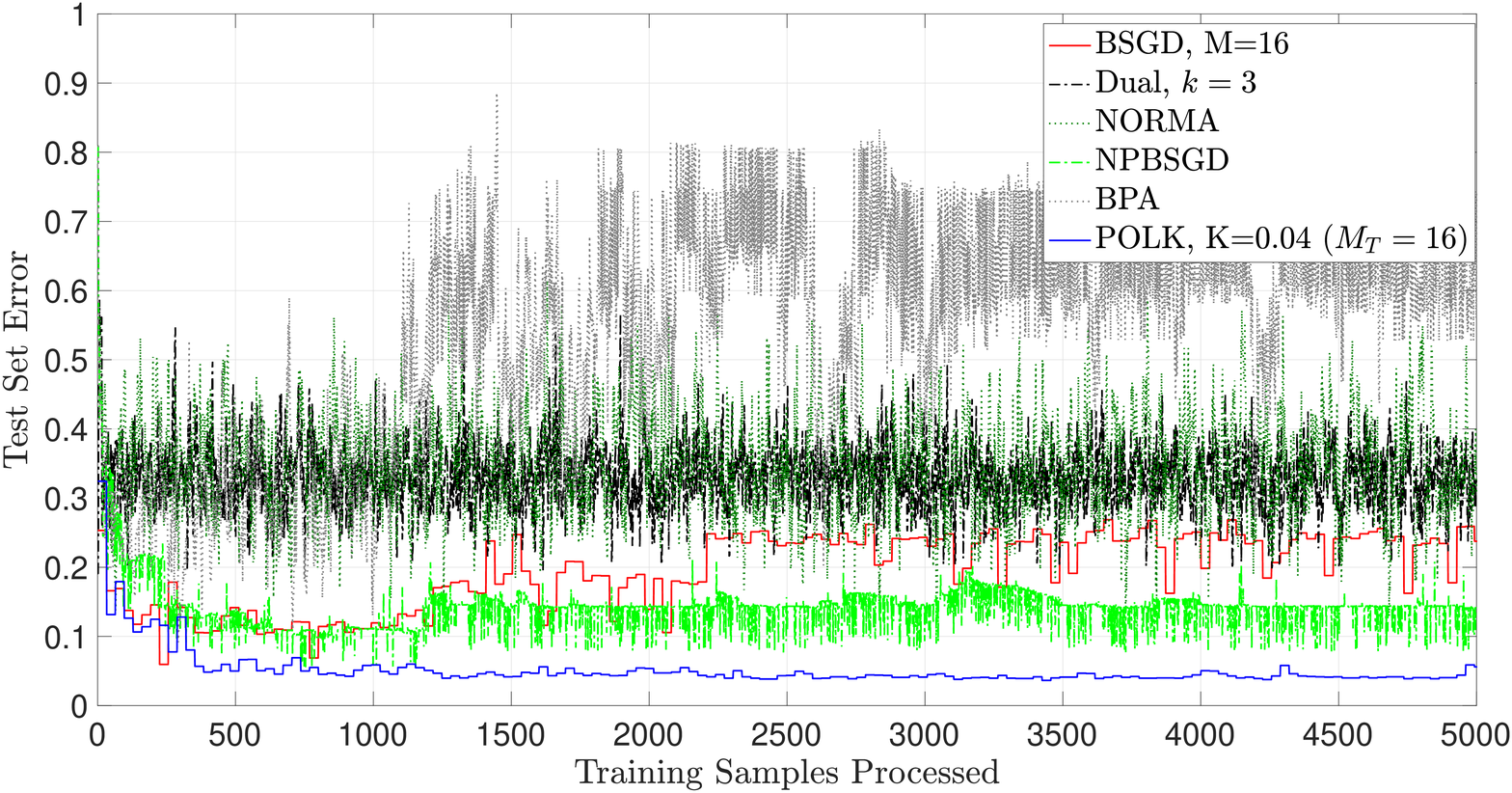}\label{subfig:error_gmm_ksvm_dense}}\hspace{-6mm}
	\subfigure[Model order $M_t$]{\includegraphics[width=0.5\linewidth,height=5cm]{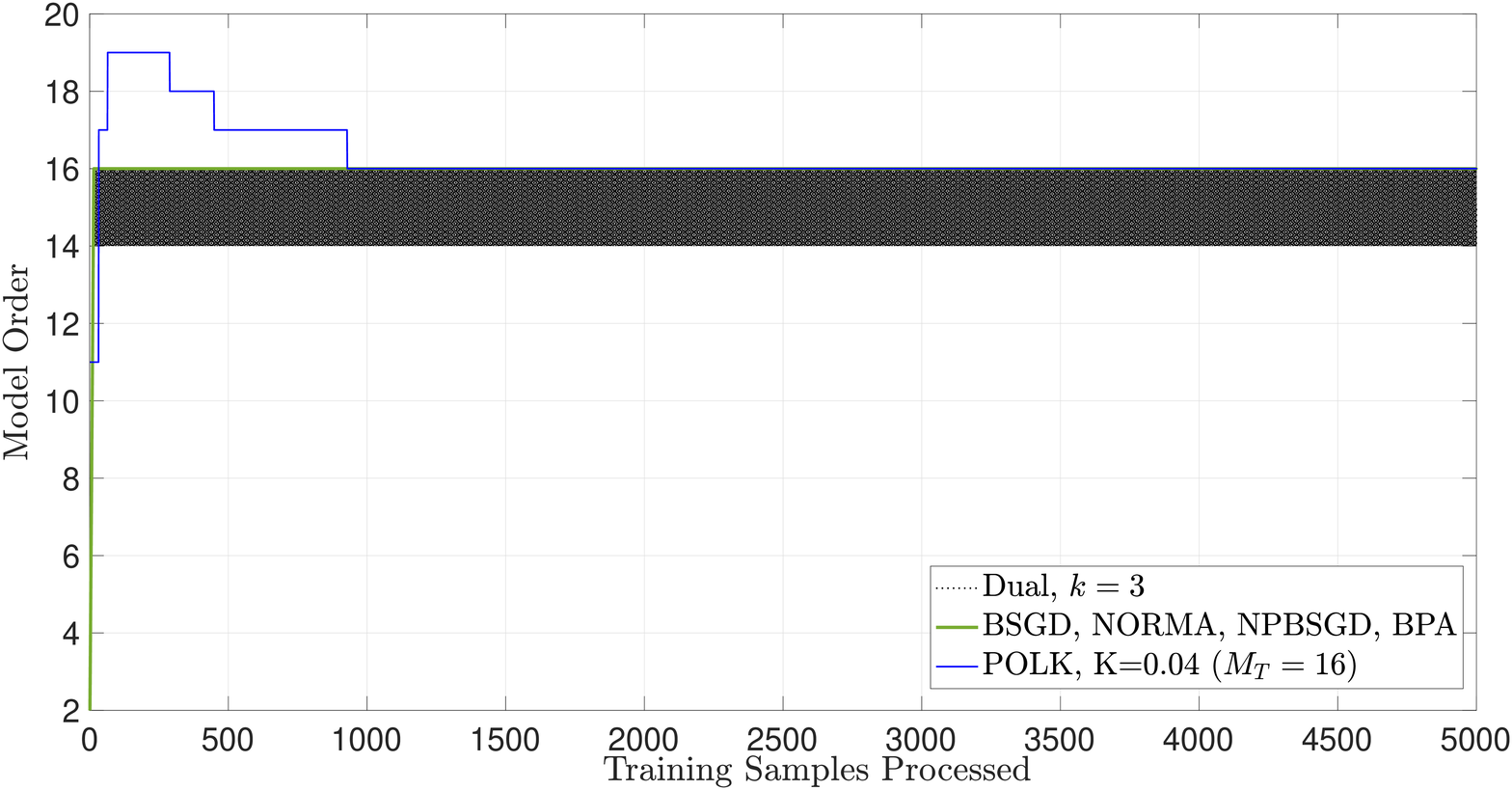}\label{subfig:order_gmm_ksvm_dense}}
	\caption{Comparison of POLK and its competitors on Gaussian Mixtures (left) for multi-class kernel SVM  when the model order parameter is fixed for the competitors at $M=16$ and the parsimony constant of POLK is set to $K=10^{-4}$. Observe that POLK achieves lower risk and higher accuracy than its competitors on this problem instance.}\vspace{-8mm}
	\label{fig:ksvm_gmm_dense}
\end{figure}

 {\bf \noindent Experiments } 
 We now present {experimental results} of Algorithm \ref{alg:soldd} for multi-class classification on Gaussian Mixtures dataset (Fig. \ref{subfig:gmm_data}) as in \cite{Zhu2005} for the case of kernel SVM, and compare with several alternatives: budgeted stochastic gradient descent (BSGD) \cite{wang2012breaking}, a fixed subspace projection method, which requires a maximum model order \emph{a priori}; {Dual Space Gradient Descent (Dual) \cite{le2016dual}, a hybrid of FSGD with a random features; nonparametric budgeted SGD (NPBSGD) \cite{le2016nonparametric}, which combines a fixed subspace projection with random dropping, Naive Online Regularized Minimization Algorithm  (NORMA) \cite{Kivinen2004}, which truncates the memory to finite-horizon objectives, and Budgeted Passive-Aggressive (BPA) which merges incoming points via nearest neighbor \cite{wang2010online}. 
 
In Figure \ref{fig:ksvm_gmm_dense} we plot the empirical results of this experiment. POLK outperforms many of its competitors by an order of magnitude in terms of test-set error rate (Fig \ref{subfig:error_gmm_ksvm_dense}). Moreover, because the marginal feature density contains $15$ modes, the optimal model order is $M^*=15$, which is approximately \emph{learned} by POLK for $K = 0.04$ (i.e., $M_T=16$) (Fig. \ref{subfig:order_gmm_ksvm_dense}). Several alternatives initialized with this parameter, on the other hand, do not converge. Moreover, POLK favorably trades off accuracy and sample complexity  -- reaching below $4\%$ error after only $1249$ samples. The final decision surface $\bbf_T$ of this trial of POLK is shown in Fig. \ref{subfig:decision_svm}, where it can be seen that the selected kernel dictionary elements concentrate at modes of the class-conditional density.  
Next, we discuss modifications of Algorithm \ref{alg:soldd} that avoid overfitting via notions of risk.

\subsection{Compositional and Risk-Aware Learning with Kernels}\label{sec:colk}

%
In this section, we explain how augmentations of \eqref{eq:kernel_stoch_opt} may incorporate risk-awareness into learning, motivated by bias-variance tradeoffs. In particular, given a particular sample path of data, one may learn a model overly sensitive to the peculiarities of the available observations, a phenomenon known as overfitting. {Overfitting has  deleterious effects when the signal-to-noise ratio of observations drops below a critical threshold, as the model will memorize noise. This occurs, for instance, in wireless communications  in complex propagation environments, object recognition with occlusions, and stock valuation during periods of high volatility. {We present concrete use cases in Figure \ref{fig:multidist_timeseries} for regression and classification on heavy-tailed training data.}}

To avoid overfitting \emph{offline}, bootstrapping (data augmentation), cross-validation, or sparsity-promoting penalties are effective \cite{friedman2001elements}. However, these approaches do not apply to streaming settings. For online problems, one must augment the objective itself to incorporate uncertainty, a topic extensively studied in econometrics \cite{levy1979approximating}. Specifically, one may use \emph{coherent risk} as a surrogate for error variance \cite{artzner1999coherent}, which permits the derivations of online algorithms that do not overfit, and are attuned to distributions with ill-conditioning or heavy-tails, as in interference channels or visual inference with less-than-laboratory levels of cleanliness.

To clarify the motivation for risk-aware augmentations of \eqref{eq:kernel_stoch_opt}, we first briefly review the bias-variance (estimation-approximation) tradeoff. Suppose we run some algorithm, {such as Algorithm \ref{alg:soldd}}, and obtain estimator $\hat f$. Then one would like to make the performance of $\hat f$ approach the Bayes optimal $\hat{\bby}^\star=\argmin_{\hat{\bby}\in\ccalY^{\ccalX}} \mathbb{E}_{\bbx, \bby}[\ell(\hat{\bby}(\bbx),\bby)] $
%
where $\ccalY^{\ccalX}$ denotes the space of \emph{all functions}  $\hat{\bby}:\ccalX\rightarrow \ccalY$ that map data $\bbx$ to target variables $\bby$. 
The performance gap between $\hat f$ and $\hat{\bby}^\star$ decomposes as \vspace{-1mm}
\begin{align}\label{eq:bias_variance}
\underbrace{\mathbb{E}_{\bbx, \bby}[\ell(\hat{f}(\bbx),\bby)]  - \min_{f\in\ccalH} \mathbb{E}_{\bbx, \bby}[\ell(f(\bbx),\bby)]}_{\text{Bias}} + 
\underbrace{\min_{f\in\ccalH} \mathbb{E}_{\bbx, \bby}[\ell(f(\bbx),\bby)] - \min_{\hat{\bby}\in\ccalY^{\ccalX}} \mathbb{E}_{\bbx, \bby}[\ell(\hat{\bby}(\bbx),\bby)]}_{\text{Variance}}
\end{align}
by adding and subtracting \eqref{eq:kernel_stoch_opt} (ignoring regularization) \cite{friedman2001elements}. 
%
\begin{figure}[t]
	\centering 
	\subfigure[Training Data]{\includegraphics[width=0.38\linewidth,height=3cm]{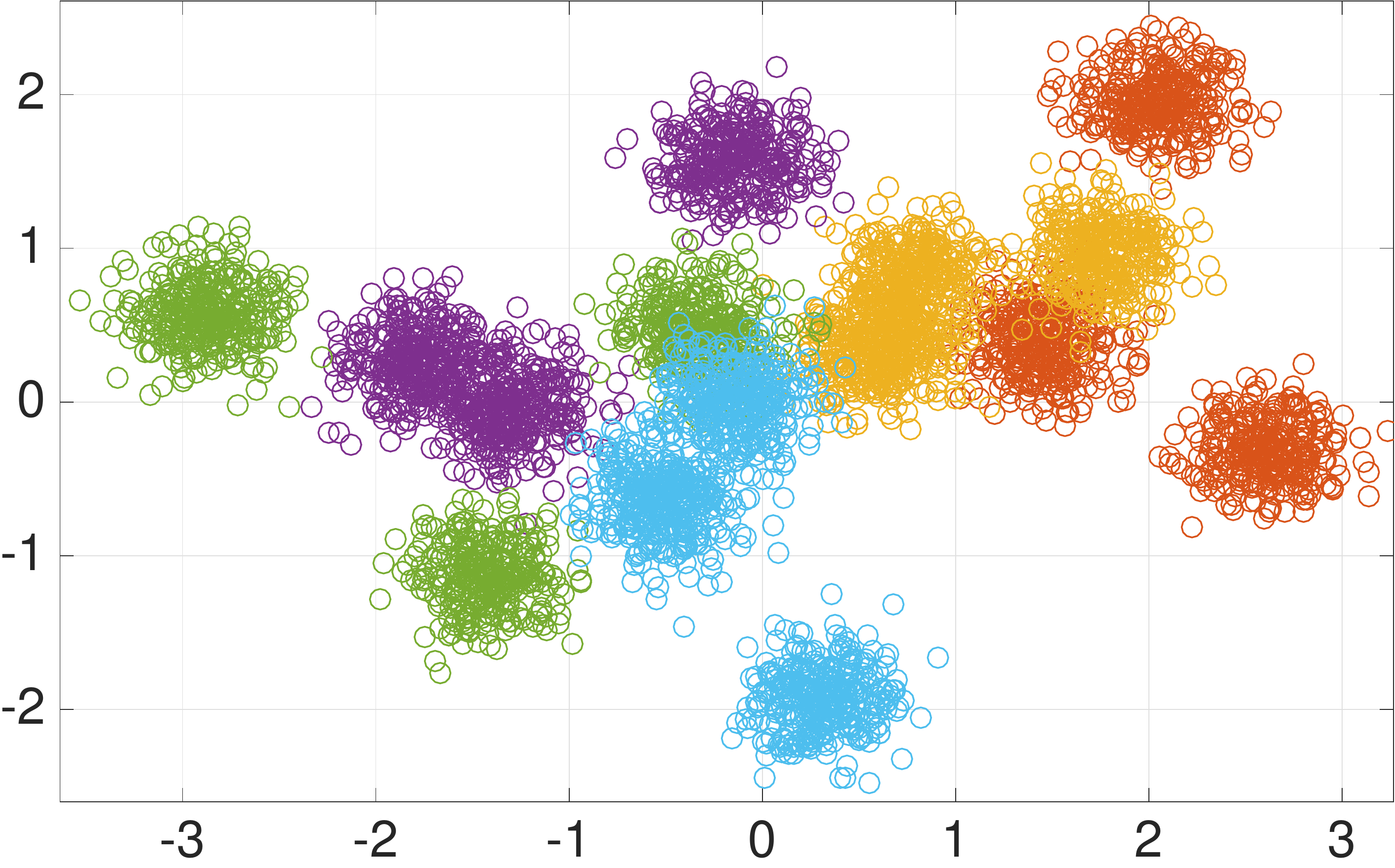}\label{subfig:gmm_data}}\hspace{10mm}
	\subfigure[$\bbf_T$ (hinge loss)]{\includegraphics[width=0.38\linewidth,height=3cm]{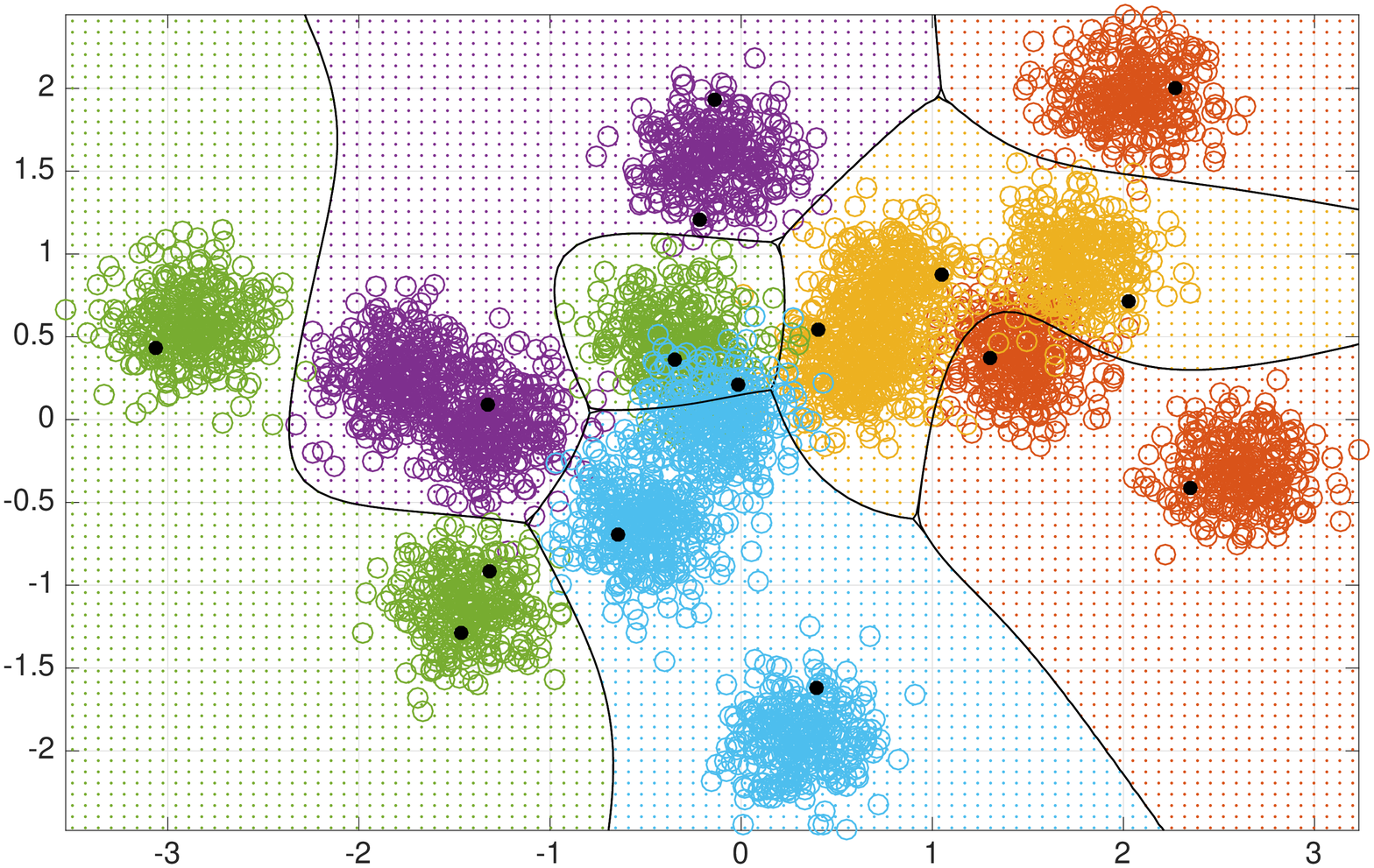}\label{subfig:decision_svm}} 
	%
	\caption{Visualization of the decision surfaces yielded by POLK for the multi-class kernel SVM and logisitic regression on Gaussian Mixtures. Training examples from distinct classes are assigned a unique color. Grid colors represent the classification decision by $\bbf_T$. Bold black dots are kernel dictionary elements, which concentrate at the modes of the data distribution. Solid lines denote class label boundaries and grid colors denote classification decisions.}
	\label{fig:decision}\vspace{-4mm}
\end{figure}
Thus, the discrepancy decomposes into two terms: the estimation error, or bias, and approximation error, or variance\footnote{Approximation error is more general than variance, but for, e.g., the quadratic loss, the former reduces into the later plus the noise of the data distribution. We conflate these quantities for ease of explanation, but they are different.}. The bias is minimized as the number of data points goes to infinity. On the other hand, universality implies the variance is null, but in practice due to inherent unknown properties of data and hyperparameter choices, it is positive.
%
%
%
To avoid overfitting in the online setting, we account for error variance via a dispersion measure $\mathbb{D}[\ell(f(\bbx),\bby)]$, thus accounting for approximation error\vspace{-1mm}
\begin{equation}\label{eq:robust_supervised_loss}
f^\star=\argmin_{f\in\ccalH} \mathbb{E}_{\bbx, \bby}[\ell(f(\bbx),\bby)]  + \mu  \mathbb{D}[\ell(f(\bbx),\bby)]
\end{equation}
%
For example, $\mathbb{D}[\ell(f(\bbx),\bby)]=
%
%
\widetilde{\text{Var}}[\ell(f\!(\bbx),\!\bby)]\!= \mathbb{E}_{\bbx, \bby}\Big\{ \!\!\big(\ell(f\!(\bbx),\!\bby)- {\mathbb{E}_{\bbx', \bby'}[\ell(f\!(\bbx'),\!\bby')]} \big)_+^2\!\!\Big\}$ 
%
%
is commonly used -- the semivariance. {Here $a_+ = \max(a,0)$ denotes the positive projection.} Alternatives are $p$-th order semideviation or the conditional value-at-risk (CVaR), which quantifies the loss function at tail-end quantiles of its distribution.

Choice of $\mu$  scales the emphasis on bias or variance error in \eqref{eq:robust_supervised_loss}, and its solutions, as compared with \eqref{eq:kernel_stoch_opt}, are attuned to outliers and higher-order moments of the data distribution. Thus, for classification, $f$ may be equipped for classification with significant class overlap, or regression (nonlinear filtering) with dips in signal to noise ratio.
%
To derive solutions to \eqref{eq:robust_supervised_loss}, we begin by noting that this is a special case of compositional stochastic programming  \cite{wang2017stochastic}, given as\vspace{-2mm}
 \begin{align}\label{eq:main_problem}
\min_{f\in\ccalH} \mathbb{E}_{\bbtheta}\left[ \ell_{\bbtheta}\left( \mathbb{E}_{\bbxi}\left[ \mathbf{\ayche}_{\bbxi}( f(\bbxi)) \right]\right) \right] + \frac{\lambda}{2}\|f \|^2_{\ccalH} \; .
\end{align}
Due to nested expectations, SGD no longer applies, and hence alternate tools are required, namely, stochastic quasi-gradient methods (SQG). Recently the behavior of SQG has been characterized in detail \cite{wang2017stochastic} -- see references therein. Here we spotlight the use of such tools for nonparametric learning by {generalizing these approaches to RKHS}, and applying matching pursuit-based projections \cite{Vincent2002}. Such an approach is the focus of \cite{bedi2019nonparametric}, which provides a tunable tradeoff between convergence accuracy and required memory, but has the additional virtue that it admits an error variance which is controllable by parameter $\mu$ in \eqref{eq:robust_supervised_loss}.

%
%
%
 We begin by writing the stochastic gradient of \eqref{eq:main_problem} as  \vspace{-2mm}
\begin{align}\label{eq:stochastic_gradient}
\langle\nabla _f\ayche_{\bbxi_t}(f(\bbxi_t)),\nabla \ell_{\bbtheta_t}(\mathbb{E}[\ayche_{\bbxi}(f(\bbxi))])\rangle+\lambda f.
\end{align}
However, the preceding expression at training examples $(\bbxi_t, \bbtheta_t)$ is not available due to the expectation involved in the argument of $ \nabla\ell\left(\cdot\right)$. A second realization of $\bbxi$ is required to estimate the inner-expectation. Instead, we use SQG, which defines a scalar sequence $\bbg_{t}$ to track the instantaneous functions $\ayche_{\bbxi_t}(f(\bbxi_t))  $ evaluated at sample pairs $\bbxi_t$:\vspace{-2mm}
\begin{align}\label{eq:auxiliary_update}
\bbg_{t+1}=(1-\beta_t)\bbg_t+\beta_t\ayche_{\bbxi_t}( f(\bbxi_t))
\end{align}
with the intent of estimating the expectation $\mathbb{E}\left[ \ayche_{\bbxi}( f(\bbxi)) \right]$. In \eqref{eq:auxiliary_update}, $\beta_t$ is a scalar learning rate chosen from the unit interval $(0,1)$ which may be either diminishing or constant. Then, we define a function sequence $f_t \in \ccalH$ initialized as null $f_0=0$, that we sequentially update using SQG:\vspace{-1mm}
\begin{align}\label{eq:sqg_descent}
\!\!\!\!f_{t+1}\!=\!(1\!-\!\lambda\alpha_t)f_t\!-\!\eta_t{\langle\nabla _f\ayche_{\bbxi_t}(f(\bbxi_t)),\nabla \ell_{\bbtheta_t}(\bbg_{t+1})\rangle}  
\!=\!(1\!-\!\lambda\alpha_t)f_t\!-\!\eta_t\langle\ayche'_{\bbxi_t}( f(\bbxi_t)),\ell'_{\bbtheta_t}(\bbg_{t+1})\rangle \kappa(\bbxi_t,\!\cdot\!) , \!\!
\end{align} 
where $\alpha_t$ is a step-size parameter chosen as diminishing or constant, and the equality makes use of the chain rule and the reproducing kernel property \eqref{eq:rkhs_properties}(i).
Through the Representer Theorem \eqref{eq:kernel_expansion}, we then have parametric updates on the coefficient vector $\bbw$ and kernel dictionary $\bbU$\vspace{-1mm}
\begin{align}\label{eq:parametric_updates}
\bbU_{t+1}=\left[\bbU_t, \bbxi_t\right] \; , \quad
\bbw_{t+1}=\left[(1-\eta_t\lambda)\bbw_t, -\eta_t\langle\ayche'_{\bbxi_t}( f(\bbxi_t)),\ell'_{\bbtheta_t}(\bbg_{t+1})\rangle\right]. 
\end{align}
In \eqref{eq:parametric_updates},  \emph{kernel dictionary} parameterizing function $f_t$ is a matrix $\bbU_t\in \reals^{p\times (t-1)}$ which stacks past realizations of $\bbxi$, and the coefficients $\bbw_t\in\reals^{t-1}$ as the associated scalars in the kernel expansion \eqref{eq:kernel_expansion} which are updated according to \eqref{eq:parametric_updates}.
The function update of \eqref{eq:sqg_descent} implies that the complexity of computing $f_t$ is $\ccalO(t)$, due to the fact that the number of columns in $\bbU_t$, or \emph{model order} $M_t$, is $(t-1)$, and thus is unsuitable for streaming settings. This computational cost is an inherent challenge of extending \cite{wang2017stochastic} to nonparametric kernel functions. To address this, one may project \eqref{eq:sqg_descent} onto low-dimensional subspaces in a manner similar to Algorithm \ref{alg:soldd} -- see \cite{koppel2019parsimonious}. The end-result, \eqref{eq:sqg_descent} operating in tandem with the projections defined in the previous section, is what we call \emph{Compositional Online Learning with Kernels} (COLK), and is summarized as Algorithm \ref{alg:colk}. Its behavior trades off convergence accuracy and memory akin to Table \ref{tab2}, and is studied in \cite{bedi2019nonparametric}.

\begin{algorithm}[t]
    \caption{Compositional Online Learning with Kernels (COLK)}        \label{alg:colk}
{    \begin{algorithmic}
        \Require $\{\bbtheta_t,\bbxi_t,\alpha_t,\beta_t,\epsilon_t \}_{t=0,1,2,...}$
        \State \textbf{initialize} ${f}_0(\cdot) = 0, \bbD_0 = [], \bbw_0 = []$, i.e. initial dictionary, coefficient vectors are empty
        \For{$t=0,1,2,\ldots$}
        \State \textbf{Update} auxiliary variable $\bbg_{t+1}$ according to \eqref{eq:auxiliary_update}\vspace{-2mm}
        \begin{align*}
        \hspace{-2.7cm} \bbg_{t+1}=(1-\beta_t)\bbg_t+\beta_t\ayche_{\bbxi_t}( f(\bbxi_t))
        \end{align*}
        \State \textbf{Compute} functional stochastic quasi-gradient step \eqref{eq:sqg_descent}\vspace{-2mm}
        \begin{align*}
        \!\!\tilde{f}_{t+1}\!=\!\ &(1-\lambda\alpha_t)f_t-\alpha_t{\langle\nabla _f\ayche_{\bbxi_t}(f(\bbxi_t)),\nabla \ell_{\bbtheta_t}(\bbg_{t+1})\rangle}  \nonumber
        \end{align*}
        \State \textbf{Revise} function parameters: dictionary $\&$ weights \eqref{eq:parametric_updates}\vspace{-2mm}
        \begin{align*}
        \tbD_{t+1} &\!= [\bbD_t,\;\;\bbxi_t \; , \bbtheta_t], \ \ \ \ 
        \tbw_{t+1} \!=\! [(1\!-\!\alpha_t\lambda)\bbw_t,\;\; \!\!-{\alpha_t\langle\nabla _f\ayche_{\bbxi_t}(f(\bbxi_t)),\nabla \ell_{\bbtheta_t}(\bbg_{t+1})\rangle}]
        \end{align*}
        \State \textbf{Compress} parameterization via KOMP
        %
        $({f}_{t+1},\bbD_{t+1},\bbw_{t+1}) = \textbf{KOMP}(\tilde{f}_{t+1},\tbD_{t+1},\tbw_{t+1},\epsilon_t)$
        
        \EndFor
    \end{algorithmic}\normalsize}
\end{algorithm}
\vspace{-0mm}


{\bf \noindent Experiments} We discuss the setup for online regression (filtering): fitness is determined by the square loss 
%
%
$\ell(f(\bbx_n),y_n) = (f(\bbx_n) - y_n)^2$
%
%
where $\bbx_n \in\ccalX\subset \reals^p$ and $y\in\reals$. Due to the bias-variance tradeoff \eqref{eq:bias_variance}, we seek to minimize both the bias and variance of the loss. {We propose accounting for the variance through a specific risk measure, namely, the $p$-th order central moment}:\vspace{-1mm}
\begin{align}\label{eq:semi_deviation}
\mathbb{D}[\ell(f\!(\bbx),\bby)]= \sum_{p=2}^P\mathbb{E}_{\bbx, \bby}\Big\{ \big(\ell(f\!(\bbx),\!\bby)\!- \!{\mathbb{E}_{\bbx', \bby'}[\ell(f(\bbx'),\!\bby')]} \big)^p\Big\} .
 \end{align}
 For the experiment purposes, we select $P=4$. We remark that the dispersion measure in \eqref{eq:semi_deviation} is non-convex which corresponds to the variance, skewness, and kurtosis of the loss distribution. We can always convexify the dispersion measure via positive projections (semi-deviations); however, for simplicity, we omit the positive projections in experiments. 
%

We evaluate COLK on data whose distribution has higher-order effects, and compare its test accuracy against existing benchmarks that minimize only bias. We inquire as to which methods overfit: COLK (Algorithm \ref{alg:colk}), as compared with BSGD \cite{wang2012breaking},  NPBSG \cite{le2016nonparametric}, POLK \cite{koppel2019parsimonious}.
 We consider $20$ different training sets from the same distribution. To generate the synthetic dataset \texttt{regression
 outliers}, we used the function $y=2x+3\text{sin}(6x)$ as the original function (a reasonable template for phase retrieval) and target $y$'s are perturbed by additive zero mean Gaussian noise. First we generate $60,000$ samples of the data, and then select $20\%$ as the test data set. From the remaining $4,800$ samples, we select $50\%$ at random to generate $20$ different training sets. We run COLK over these training sets with parameter selections: a Gaussian kernel with bandwidth $\sigma=.06$, step-size $\eta=0.02$, $\beta=0.01$, $\epsilon=K\alpha^2$ with parsimony constant $K=5$, variance coefficient $\eta=0.1$, and mini-batch size of $1$. Similarly, for POLK we use $\alpha=0.5$ and  $\epsilon=K\alpha^2$ with parsimony constant $K=0.09$. We fix the kernel type and bandwidth, and the parameters that define comparator algorithms are hand-tuned to optimize performance with the restriction that their complexity is comparable.  We run these algorithms over different training realizations and evaluate their test accuracy as well as standard deviation.
\begin{figure}
	\subfigure[Visualization of regression.]{\includegraphics[width=0.34\linewidth,height=4.2cm]{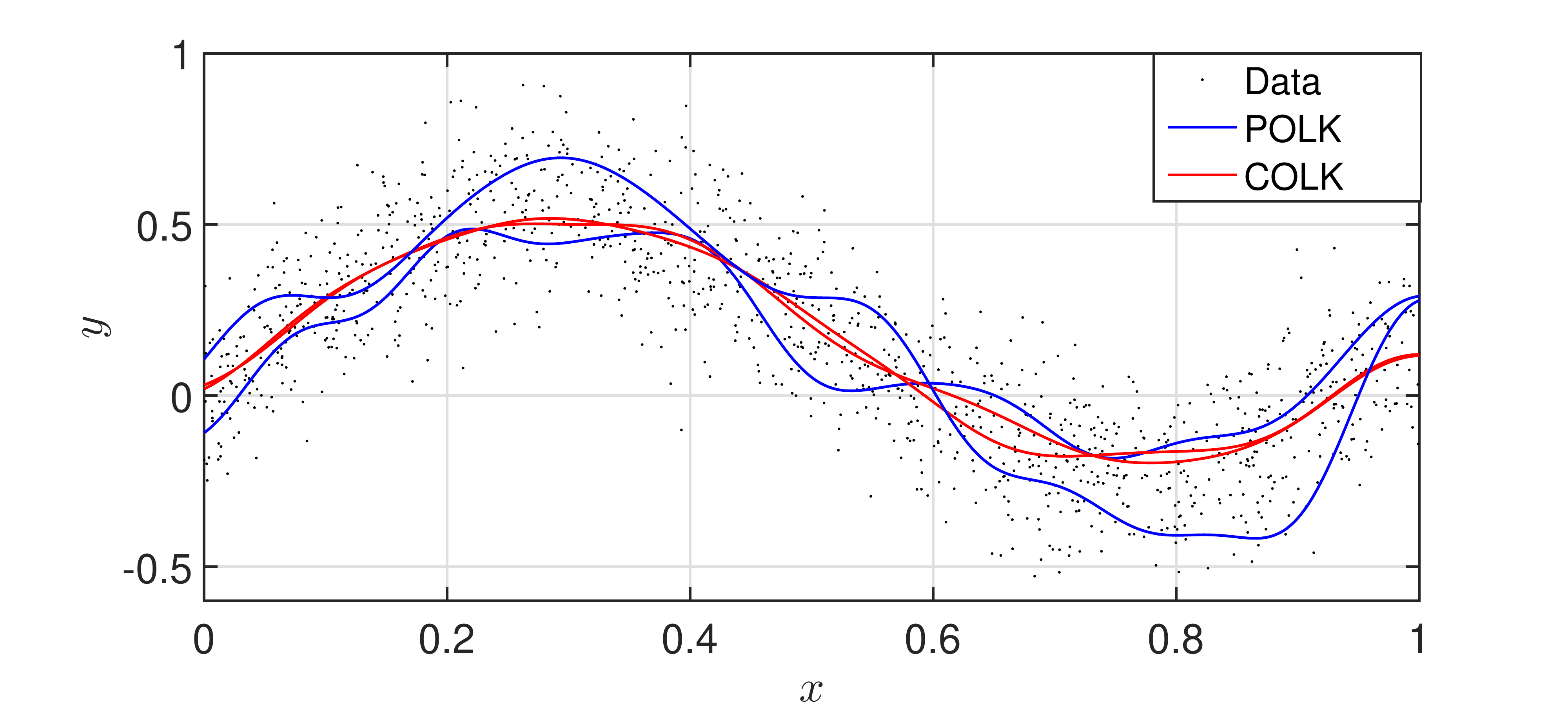}
			\label{fig:visualization}}\hspace{-5mm}	
	\subfigure[Model order]{\includegraphics[width=0.34\linewidth,height=4.2cm]{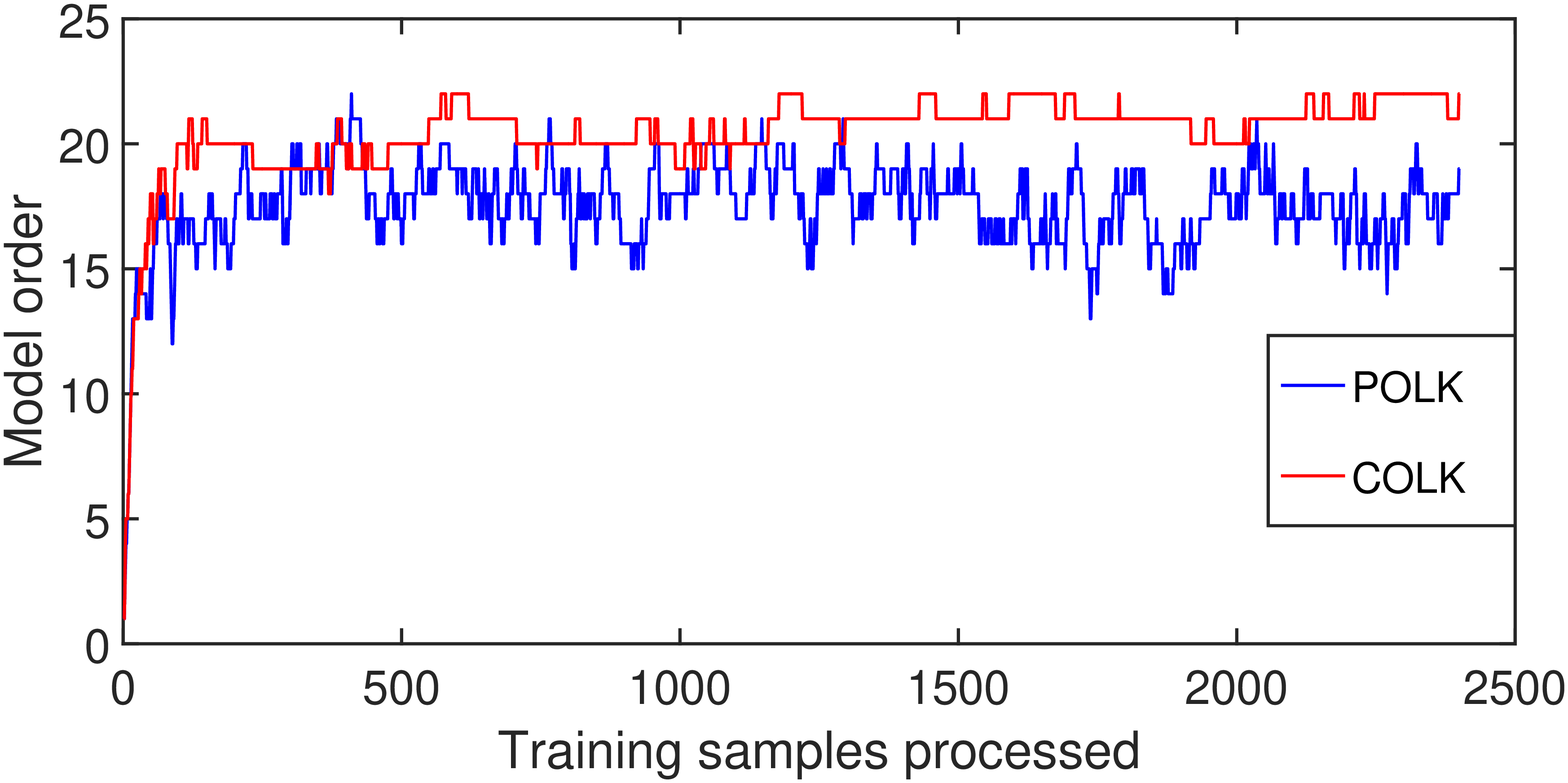}\label{fig:model_order}}	\hspace{-5mm}		
		\subfigure[Test Error and Std. dev.]{\includegraphics[width=0.34\linewidth,height=4.2cm]{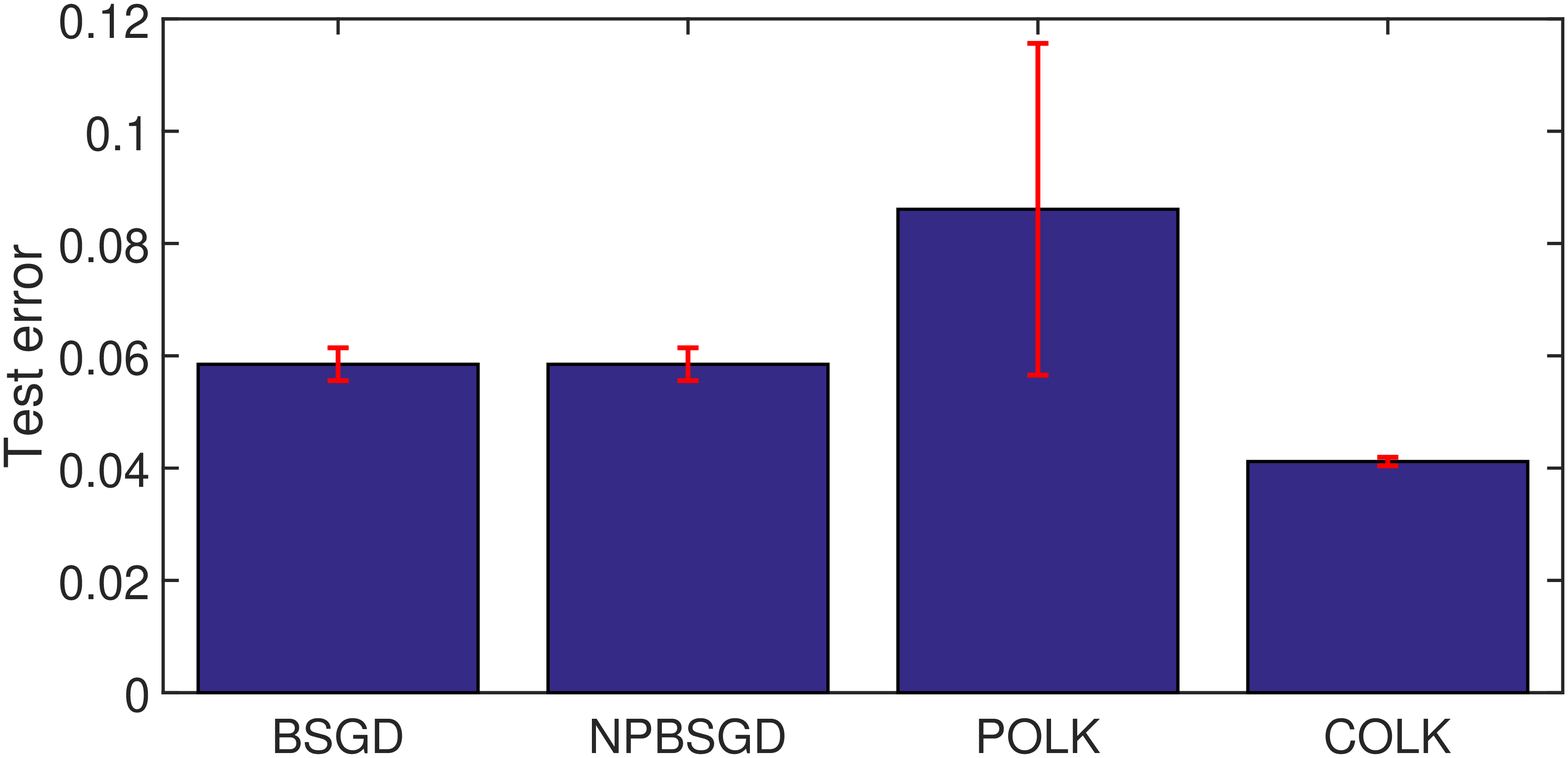}
		\label{fig:box}}
%
\vspace{-2mm}	\caption{\small COLK, with $\alpha=0.02$, $\eps=\alpha^2$, $\beta=0.01$, $K=5$, $\eta=0.1$, bandwidth $c=.06$ as compared to alterantives for online learning with kernels that only minimize bias for \texttt{regression outliers} data. COLK quantifies risk as variance, skewness, and kurtosis. We report test MSE averages over $20$ training runs, and its standard dev. as error bars. Outlier presence does not break learning stability, and test accuracy remains consistent, at the cost of increased complexity. COLK attains minimal error and deviation.
}\label{fig:multidist_timeseries}\vspace{-8mm}
\end{figure}

The advantage of minimizing the bias as well as variance may be observed in Fig.~~\ref{fig:visualization} which plots the learned function for POLK and COLK for two training data sets. POLK learning varies from one training set to anoother while COLK is robust to this change. 
In Fig.~\ref{fig:model_order} we plot the model order of the function sequence defined by COLK, and observe it stabilizes over time regardless of the presence of outliers. These results substantiate the convergence behavior spotlighted in \cite{bedi2019nonparametric}, which also contains additional experimental validation on real data.

\section{Decentralized Learning Methods}\label{sec:decentralized}
In domains such as autonomous networks of robots or smart devices, data is generated at the network edge. In order to gain the benefits of laws of large numbers in learning, aggregation of information is required. 
However, transmission of raw data over the network may not be viable or secure, motivating the need for decentralized processing. Here, the goal is for each agent, based on local observations, to learn online an estimator as good as a centralized one with access to all information in advance. To date, optimization tools for multi-agent online learning are predominately focused to cases where agents learn linear statistical models {(vector parameters)} \cite{nedic2009distributed}. However, since kernel learning may be formulated as a stochastic convex problem over a function space, standard strategies, i.e., distributed gradient \cite{koppel2018decentralized} and primal-dual methods \cite{pradhan2018exact} may be derived. Doing so is the focus of this section, leveraging the projection in Algorithm \ref{alg:soldd}.
%

To formulate decentralized learning, we consider an undirected graph $\mathcal{G}=(\ccalV, \ccalE)$ with $V = |\ccalV|$ nodes and $E = |\ccalE|$ edges. Each $i\in V$ represents an agent in the network, who observes a distinct observation sequence $(\bbx_{i,t},y_{i,t}) \sim (\bbx_{i},y_{i}) $ and quantifies merit according to their local loss $\mbE_{\bbx_i,y_i}\Big[ \ell_i(f\big(\bbx_i), y_i\big)\Big]$. Based on their local data streams, they would like to learn as well as a clairvoyant agent which has access to global information at all {times}:\vspace{-1mm}
\begin{align}\label{eq:kernel_stoch_opt_global}
\!\!\!\!f^\star:= \argmin_{f \in \ccalH}\sum_{i\in\ccalV}\left(\mbE_{\bbx_i,y_i}\Big[ \ell_i(f\big(\bbx_i). y_i\big)\Big] \right)
\end{align}
{\bf \noindent Decentralized learning with consensus constraints: } 
%
Under the hypothesis that all agents seek to learn a \emph{common} global function, i.e., agents' observations are uniformly relevant to others, one would like to solve \eqref{eq:kernel_stoch_opt_global} in a decentralized manner. To do so, we define local copies $f_i$ of the global function, and reformulate \eqref{eq:kernel_stoch_opt_global} as\vspace{-1mm}
\begin{align}\label{eq:kernel_stoch_opt_consensus}
\!\!\!\!f^\star&:= \argmin_{\{f_i \in \ccalH\}}\sum_{i\in\ccalV}\left(\mbE_{\bbx_i,y_i}\Big[ \ell_i(f_i\big(\bbx_i), y_i\big)\Big] \right) 
\qquad \text{s.t. } f_i = f_j \hspace{1cm} (i,j)\in\ccalE
\end{align}
Imposing \emph{functional} constraints of the form $f_i=f_j$ in \eqref{eq:kernel_stoch_opt_consensus} is challenging because it involves computations independent of data, and hence may operate outside the realm of the Representer Theorem \eqref{eq:kernel_expansion}. To alleviate this issue, we approximate consensus in the form $f_i(\x_i) = f_j(\x_i)$ which is imposed for $(i,j)\in\ccalE$ in expectation over $\x_i$. {Thus agents are incentivized to agree regarding their estimates without needing to agree on their functions}. This modification of consensus yields a penalty functional amenable to efficient computations, culminating local updates for each $i\in\ccalV$:
\begin{align}\label{eq:sgd_hilbert_local}
f_{i,t+1} &=(1-\eta_t \lambda ) f_{i,t} - \eta_t \Big[\ell_i'(f_{i,t}(\bbx_{i,t}),y_{i,t}) \kappa(\bbx_{i,t},\cdot) + \! c  \!\sum_{j\in n_i} ( f_{i,t}(\bbx_{i,t}) \! - \! f_{j,t}(\bbx_{i,t}))\kappa(\bbx_{i,t},\cdot) \Big] 
\end{align}
where $c$ is a penalty coefficient that ensures $\mathcal{O}(1/c)$ constraint violation, and exact consensus as $c\rightarrow \infty$. {To implement \eqref{eq:sgd_hilbert_local} in a decentralized manner, agent $i$ must send its local observation $\bbx_{i,t}$ to neighboring agents $j\in n_i$, who respond with their function evaluation $f_{j,t}(\bbx_{i,t})$. Neighboring agents $j$ respond in kind to obtain $f_{i,t}(\bbx_{j,t})$. Overall computation relative to a centralized scheme is reduced when the network is large.}
Moreover, stacking the functions $f_i$ along $i\in V$ by $f_t$, \cite{koppel2018decentralized} establishes tradeoffs between  convergence and memory akin to Table \ref{tab2} hold for decentralized learning \eqref{eq:kernel_stoch_opt_consensus} when local functions are fed into local projection steps. Experiments then establish the practical usefulness of \eqref{eq:sgd_hilbert_local} for attaining state of the art decentralized learning.

{\bf \noindent Decentralized learning with proximity constraints: } When the hypothesis that all agents seek to learn a \emph{common} global function is invalid, due to heterogeneity of agents' observations or processing capabilities, imposing consensus \eqref{eq:kernel_stoch_opt_consensus} degrades decentralized learning \cite{koppel2017proximity}.

%
%
 Thus, we define a relaxation of consensus \eqref{eq:kernel_stoch_opt_consensus} using proximity constraints that incentivizes coordination without requiring agents' decisions to coincide:\vspace{-1mm}
\begin{align}\label{eq:kernel_stoch_opt_prox}
\!\!\{f_i^\star\!\}\!:= \argmin_{\{f_i \in \ccalH\}}\sum_{i\in\ccalV}\!\!\left(\mbE_{\bbx_i,y_i}\Big[ \ell_i(f_i\big(\bbx_i), y_i\big)\Big] \right) 
\quad \text{s.t. } \mbE_{\bbx_i}\!\!\left[h_{ij}(f_{i,t}(\bbx_{i,t}), f_{j,t}(\bbx_{i,t}))\right] \leq \gamma_{ij} \quad (i,j)\in\ccalE
\end{align}
where $h_{ij}(f_{i,t}(\bbx_{i,t}), f_{j,t}(\bbx_{i,t}))$ is small when $f_{i,t}(\bbx_{i,t})$ and $f_{j,t}(\bbx_{i,t})$ are close, and $\gamma_{ij}$ defines a tolerance. This allows local solutions of \eqref{eq:kernel_stoch_opt_prox} to be different at each node, and for instance, to incorporate correlation priors into algorithm design. To solve \eqref{eq:kernel_stoch_opt_prox}, we propose a method based on Lagrangian relaxation, specifically, a functional  stochastic variant of the Arrow-Hurwicz primal-dual (saddle point) method \cite{pradhan2018exact}. Its specific form is given as follows:
\begin{align}\label{eq:primal_kernel}
f_{i,t+1} &= (1-\eta\lambda)f_{i,t} - \eta\left[\ell_i'(f_{i,t}(\bbx_{i,t}),y_{i,t}) + \sum_{j:(i,j)\in\ccalE} \mu_{ij}h'_{ij}(f_{i,t}(\bbx_ 
{i,t}), f_{j,t}(\bbx_{i,t}))\right]\kappa(\x_{i,t},\cdot)\\
\label{eq:dual_kernel}
\mu_{ij,t+1} &= \left[(1-\delta\eta^2)\mu_{ij,t} + \eta(h_{ij}(f_{i,t}(\bbx_ 
{i,t}), f_{j,t}(\bbx_{i,t}))-\gamma_{ij})\right]_{+}.
\end{align}
{To implement \eqref{eq:primal_kernel} - \eqref{eq:dual_kernel}, a comparable message passing scheme to that which is detailed following \eqref{eq:sgd_hilbert_local} is required. The key difference for the primal-dual method is that scalar dual variables $\mu_{ij, t}$ must also be sent from node $i$ to node $j$.} The KOMP-based projection is applied to each local primal update \eqref{eq:primal_kernel}, which permits us to trade off accuracy and model complexity. 

{The algorithms given here are functions of random processes, and hence their convergence is defined probabilistically. We focus on expected performance since obtaining a strict Lyapunov function for \eqref{eq:primal_kernel}-\eqref{eq:dual_kernel} remains elusive in the literature, and is required for stronger (almost sure) forms of convergence.}
Thus, define $S(f_t):=\sum_{i\in\ccalV}\E{\ell_i(f_{i,t}(\x_{i,t}),y_{i,t})} + \frac{\lambda}{2}\norm{f_{i,t}}^2_\ccalH$ as the regularized penalty. Then for horizon $T$ and step-size $\eta = \ccalO(1/\sqrt{T})$ and budget $\epsilon = \ccalO(1/T)$, we obtain sublinear growth of the sub-optimality and constraint violation as \vspace{-1mm}
\begin{align}
\sum_{t=1}^T \E{S(f_t)} \!\!- \!\!S(f^\star) \leq \ccalO(T^{1/2})\; , \quad \label{proxog}
\sum_{(i,j)\in\ccalE} \E{\sum_{t=1}^T (h_{ij}(f_{i,t}(\bbx_ 
	{i,t}), f_{j,t}(\bbx_{i,t}))-\gamma_{ij})}_{+}\!\! \leq \ccalO(T^{3/4}). 
\end{align}
Note the quantities on the right of \eqref{proxog} aggregate terms obtained over $T$ iterations, but are still bounded by sublinear functions of $T$. In other words, the average optimality gap and constraint violation are respectively bounded by $\ccalO(T^{-{1}/{2}})$ and $\ccalO(T^{-{1}/{4}})$, and approach zero for large $T$. In \cite{pradhan2018exact}, the experimental merit of \eqref{eq:primal_kernel}-\eqref{eq:dual_kernel} is demonstrated for decentralized online problems where nonlinearity is inherent to the observation model.

\section{Discussion and Open Problems}\label{sec:conclusion}
%
Algorithm \ref{alg:soldd} yields nearly optimal online solutions to nonparametric learning (Sec. \ref{sec:polk}), while ensuring the memory never becomes unwieldy. Several open problems may be identified as a result, such as, e.g., the selection of kernel hyper-parameters to further optimize performance, of which a special case has recently been studied \cite{peifer2019sparse}. Moreover, \emph{time-varying} problems where the decision variable is a function, as in trajectory optimization, remains unaddressed. On the practical side, algorithms developed in this paper may be used for, e.g., online occupancy mapping-based localization amongst obstacles, dynamic phase retrieval, and beamforming in mobile networks.  

The use of risk measures to overcome online overfitting may be used to attain online algorithms that are robust to unpredictable environmental effects (Sec. \ref{sec:colk}),  which is an ongoing challenge in indoor and urban localization \cite{elvira2019multiple}, as well as model mismatch in autonomous control \cite{koller2018learning}. Their use more widely in machine learning may reduce the ``brittleness" of deep learning as well.

For decentralized learning, numerous enhancements of the methods in Sec. \ref{sec:decentralized} are possible, such as those which relax conditions on the network, the smoothness required for stability, incorporation of agents' ability to customize hyper-parameters to local observations, and reductions of communications burden, to name a few. Online multi-agent learning with nonlinear models may pave the pathway for next-generation distributed intelligence.

The general principle of sparsifying a nonparametric learning algorithm as much as possible while ensuring a descent-like property also holds when one changes the metric, ambient space, and choice of learning update rule, as has been recently demonstrated for Gaussian Processes \cite{koppelconsistent}. Similar approaches are possible for Monte Carlo methods \cite{elvira2017adapting}, and it is an open question which other statistical methods limited by the curse of dimensionality may be gracefully brought into the memory-efficient online setting through this perspective.

Overall, the methods discussed in this work possess a conceptual similarity to  rate distortion theoretical results in the context of lossy data compression. In particular, to estimate a (class-conditional or regression) probability density with some fixed bias, one only needs finitely many points, after which all additional training examples are redundant. Such a phenomenon may be used to employ nonparametric methods in streaming problems for future learning systems.

\bibliographystyle{IEEEtran}
\bibliography{bibliography} 

%

   \end{document}